\documentclass[3p,times]{elsarticle}

\usepackage{ecrc}
\usepackage{caption}
\usepackage{subcaption}
\usepackage[rightcaption]{sidecap}

\volume{00}

\firstpage{1}

\journalname{Nuclear Physics A}

\runauth{Misha Veldhoen}

\jid{nupha}

\usepackage{amssymb}

\usepackage[figuresright]{rotating}

\begin{document}

\begin{frontmatter}



\dochead{}

\title{$p/\pi$ Ratio in Di-Hadron Correlations}


\author[uu]{Misha Veldhoen for the ALICE collaboration}
\ead{m.veldhoen@uu.nl}
\address[uu]{Institute for Subatomic Physics, Utrecht University, 3584CC Utrecht, The Netherlands}

\begin{abstract}
Particle ratios are important observables used to constrain models of particle production in heavy-ion collisions. In this work we report on a measurement of the $p/\pi$ ratio in the transverse momentum range $2.0 < p_{T,assoc} < 4.0$ GeV/c, associated with a charged trigger particle of $5.0 < p_{T,trig} < 10.0$ GeV/c, in 0-10\% central Pb--Pb collisions at $\sqrt{s_{NN}}=2.76$ TeV. The ratio is measured in the jet peak and in a region at large $\Delta\eta$ separation from the peak (bulk region). The presented results are based on 14M minimum-bias Pb--Pb collisions, recorded by the ALICE detector. It is observed that the $p/\pi$ ratio in the bulk region is compatible with the $p/\pi$ ratio of an inclusive measurement, and is much larger than the $p/\pi$ ratio in the jet peak. The $p/\pi$ ratio in the jet peak is compatible with a PYTHIA reference, in which fragmentation in the vacuum is the dominant mechanism of particle production.
\end{abstract}


\end{frontmatter}


\section{Introduction}
An important open issue in heavy-ion physics is the mechanism of particle production.  In a heavy-ion collision it is common to make a distinction between the thermal medium, produced by soft processes, and hard partons, which traverse through the medium, and form a jet. 

To study the particle production in the medium, inclusive baryon-to-meson ratios have been measured, both at RHIC and LHC. These measurements have revealed a significant enhancement of the baryon-to-meson ratio at intermediate transverse momentum ($p_T$), compared to a proton-proton reference \cite{IdentifiedHadronsPHENIX2003, IdentifiedHadronsSTAR2010Pub,IdentifiedHadronsALICE2011}. This indicates that the dominant mechanism of particle production in the thermal medium is not vacuum fragmentation. At $p_T < 2.0$ GeV/c, hydrodynamic models have been able to reproduce this enhancement, and at $2.0 <  p_T < 4.0$ GeV/c, recombination models were successful \cite{FirstResultsPbPbAtLHC2012}. Based on model calculations, some authors have suggested that also the particle production in jets in heavy-ion collisions is modified compared to the vacuum case, because of the presence of the medium \cite{JetHadrochemistry2008, CoalescenceModels2008, PartonRecombination2009, DiHadronCorrInJets2004}. 

So far experiments have shown that some jet properties seem to be modified by the presence of the medium, while others are not. For example, the shape of the near-side peak in a 2-particle correlation with a high-$p_T$ trigger particle seems to be significantly modified \cite{JanFieteJetShapes2012}. On the other hand, recent results from CMS have shown that the jet fragmentation function in central Pb--Pb collisions does not show any modification compared to the jet fragmentation function in p--p collisions. Interestingly, this seems to hold not only for the near-side jet, but also for the suppressed away-side jet \cite{CMSJetFragmentation2012}.
\nopagebreak

In this work we present an extension on the inclusive measurement of the $(p + \bar{p})/(\pi^+ + \pi^-)$ ratio, which will hereafter be referred to as $p/\pi$ ratio. By measuring the $p/\pi$ ratio in the particle yield associated with a high-$p_T$ trigger particle, a distinction can be made between the $p/\pi$ ratio in particles coming from the medium, and in particles in the jet. We will present a measurement of the $p/\pi$ ratio in the jet and in a region at large $\Delta\eta$ separation from the peak (bulk region), and make a comparison with PYTHIA, where particle production is dominated by vacuum fragmentation.

\begin{figure}
\centering
	\begin{subfigure}[b]{0.42\textwidth}
		\centering
		\includegraphics[width=\textwidth]{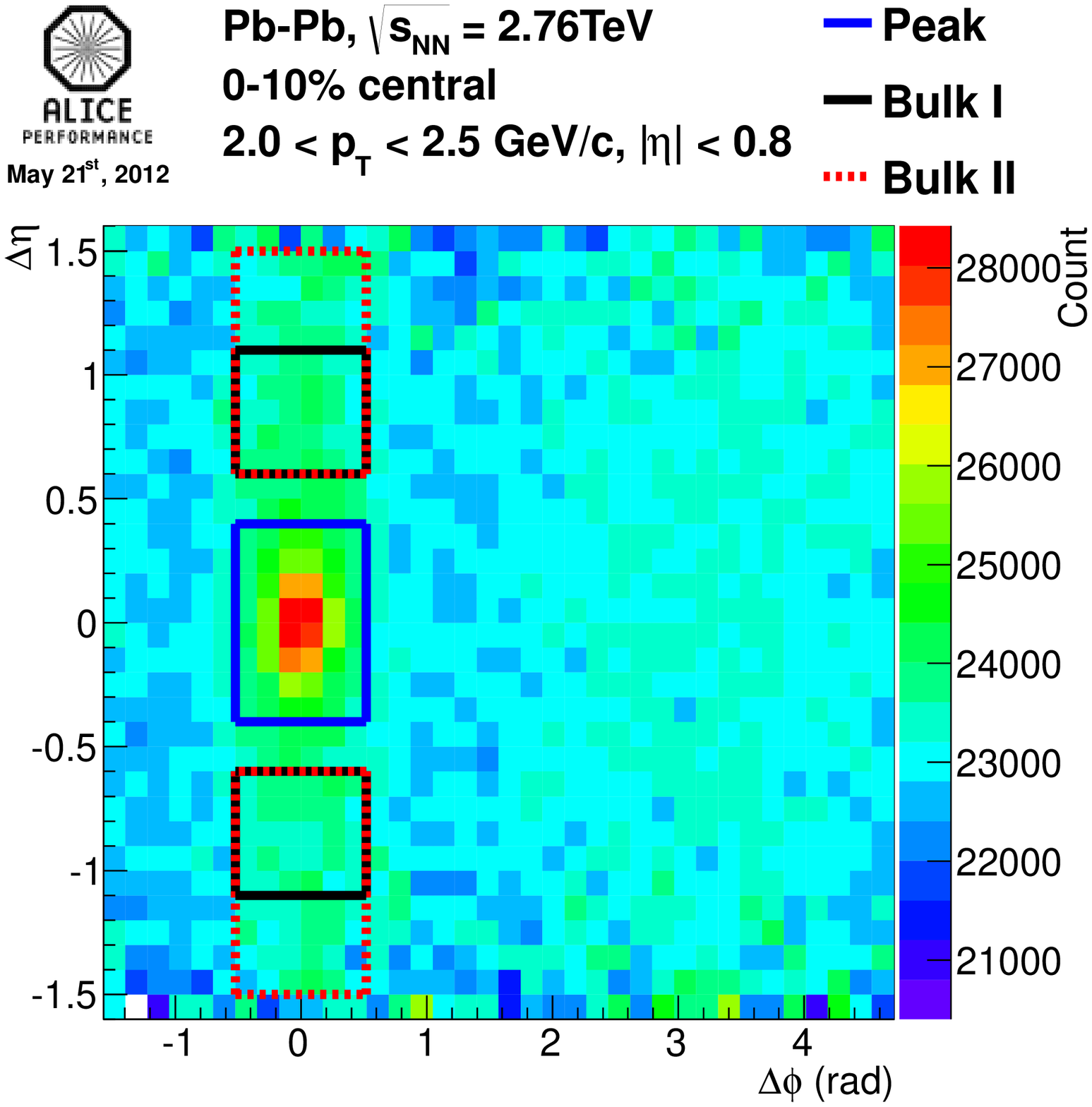}
	\end{subfigure}
	\begin{subfigure}[b]{0.42\textwidth}
		\centering
		\includegraphics[width=\textwidth]{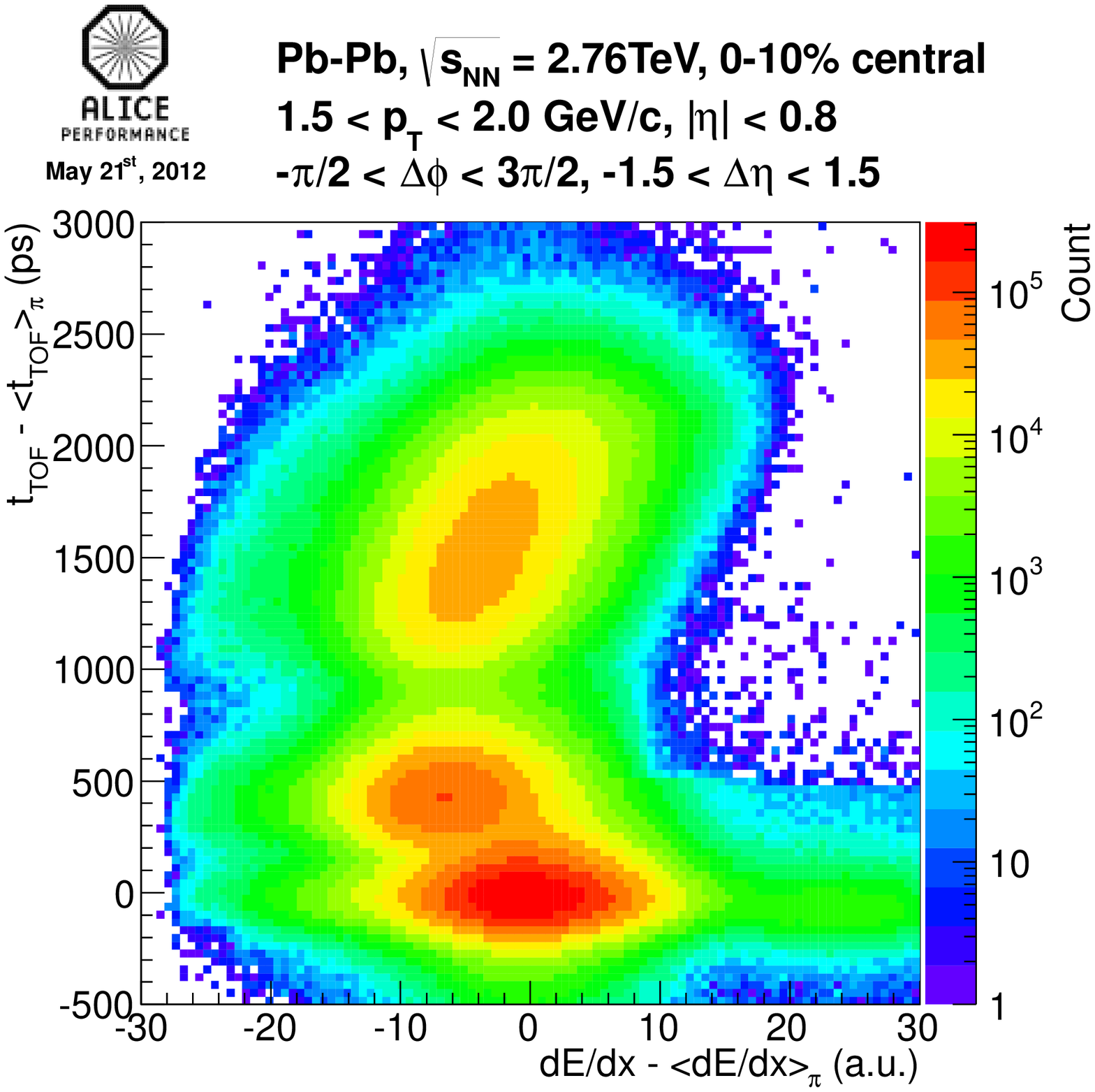}
	\end{subfigure}
	\caption{Left panel: definition of the regions in the $(\Delta\phi,\Delta\eta)$ plane. Right panel: example of a distribution of associated tracks in the ($dE/dx - \langle dE/dx\rangle_i,\Delta t - \langle\Delta t\rangle_i$) plane, under a pion mass assumption. The three most prominent peaks correspond from top to bottom to: protons, kaons and pions.}\label{fig:Projections}
\end{figure}

\section{Analysis}
For the analysis 1.4M 0-10\% central minimum-bias Pb--Pb events were used, recorded by the ALICE detector. Out of this a subset of events containing at least one high-$p_T$ trigger particle ($5.0 < p_{T,trig} <10.0$ GeV/c) was selected. For every trigger particle, the associated hadron distribution was measured at intermediate $p_T$ ($1.5 < p_{T,assoc} < 4.5$ GeV/c). For every associated track the specific energy loss ($dE/dx$), as well as the Time Of Flight ($\Delta t$) was measured. All relevant information was collected in two 5-dimensional histograms:
\begin{equation}\label{eq:FiveDimHisto}
C_i\left(\Delta\phi,\Delta\eta,p_{T,assoc},\frac{dE}{dx} - \Big\langle\frac{dE}{dx}\Big\rangle_i, \Delta t - \langle\Delta t\rangle_i\right),\qquad i\in\left\{\pi,p\right\},
\end{equation}
where $\Delta\phi \equiv \phi_{trig} - \phi_{assoc}$, $\Delta\eta \equiv \eta_{trig}-\eta_{assoc}$, and $\langle\rangle_i$ stands for the expected value for particle species $i$. The associated proton or pion yield in a certain region in $(\Delta\phi,\Delta\eta)$ and $p_{T,assoc}$ can be measured by integrating the corresponding 5-dimensional histogram in Eq. (\ref{eq:FiveDimHisto}) over the first three coordinates, and then fitting the resulting 2-dimensional histogram in $(dE/dx - \langle dE/dx\rangle_i,\Delta t - \langle\Delta t\rangle_i)$ (see for example the right panel of Fig. \ref{fig:Projections}). 

The width of the peak in the $(dE/dx - \langle dE/dx\rangle_i,\Delta t - \langle\Delta t\rangle_i)$-plane corresponding to the mass-assumption is determined by the detector resolution, and is therefore well-described by a two-dimensional Gaussian with an exponential tail in the positive $\Delta t$ direction. Since in general $\langle dE/dx\rangle_i - \langle dE/dx\rangle_j$ and $\langle\Delta t\rangle_i - \langle\Delta t\rangle_j$ for $i\neq j$ are non-trivial functions of $p_T$ and $\eta$, the peaks which do not correspond to the mass-assumption will have very different shapes (Fig. \ref{fig:Projections}). It is not easy to find an analytic description of the shape of the peaks which do not correspond to the mass-assumption. Therefore a toy Monte Carlo model was set up to generate template histograms resembling these shapes. 

Using the analytic description for the peak corresponding to the mass-assumption, and generated template histograms for the other two peaks, the histograms in the $(dE/dx - \langle dE/dx\rangle_i,\Delta t - \langle\Delta t\rangle_i)$-plane were fitted, and the identified associated yield per trigger particle was extracted as a function of $p_{T,assoc}$ for three different regions in $(\Delta\phi,\Delta\eta)$. These three regions are defined as (see the left panel of Fig. \ref{fig:Projections}):
\begin{itemize}
\item Peak region ($- 0.52 < \Delta\phi < 0.52$ rad., and $-0.40 < \Delta\eta < 0.40$),
\item Bulk I region ($- 0.52 < \Delta\phi < 0.52$ rad., and $0.60 < |\Delta\eta| < 1.10$),
\item Bulk II region ($- 0.52 < \Delta\phi < 0.52$ rad., and $0.60 < |\Delta\eta| < 1.50$).
\end{itemize}
The yields in the ``Bulk II" region were used as bulk measurement, and will be referred to as ``Bulk". The region ``Bulk I"  was used to estimate the systematic effect of changing the bulk region.
Using the measured yields $N_{i,J}$, $i\in\{\pi,p\}$, $J\in\{Peak,Bulk\}$, the ratios in the jet and in the bulk can be calculated:
\begin{equation}\label{eq:RatioDef}
R^{p/\pi}_{Bulk} = \frac{N_{p,Bulk}}{N_{\pi,Bulk}},\qquad R^{p/\pi}_{Jet} = \frac{N_{p,Peak}-(A_{Peak}/A_{Bulk}) N_{p,Bulk}}{N_{\pi,Peak}-(A_{Peak}/A_{Bulk}) N_{\pi,Bulk}},
\end{equation}
where $A_{J}$, $J\in\{Peak,Bulk\}$ stands for the area of region $J$. 

\section{Results}

 \begin{figure}
 \centering
	\begin{subfigure}[b]{0.39\textwidth}
		\centering
		\includegraphics[width=\textwidth]{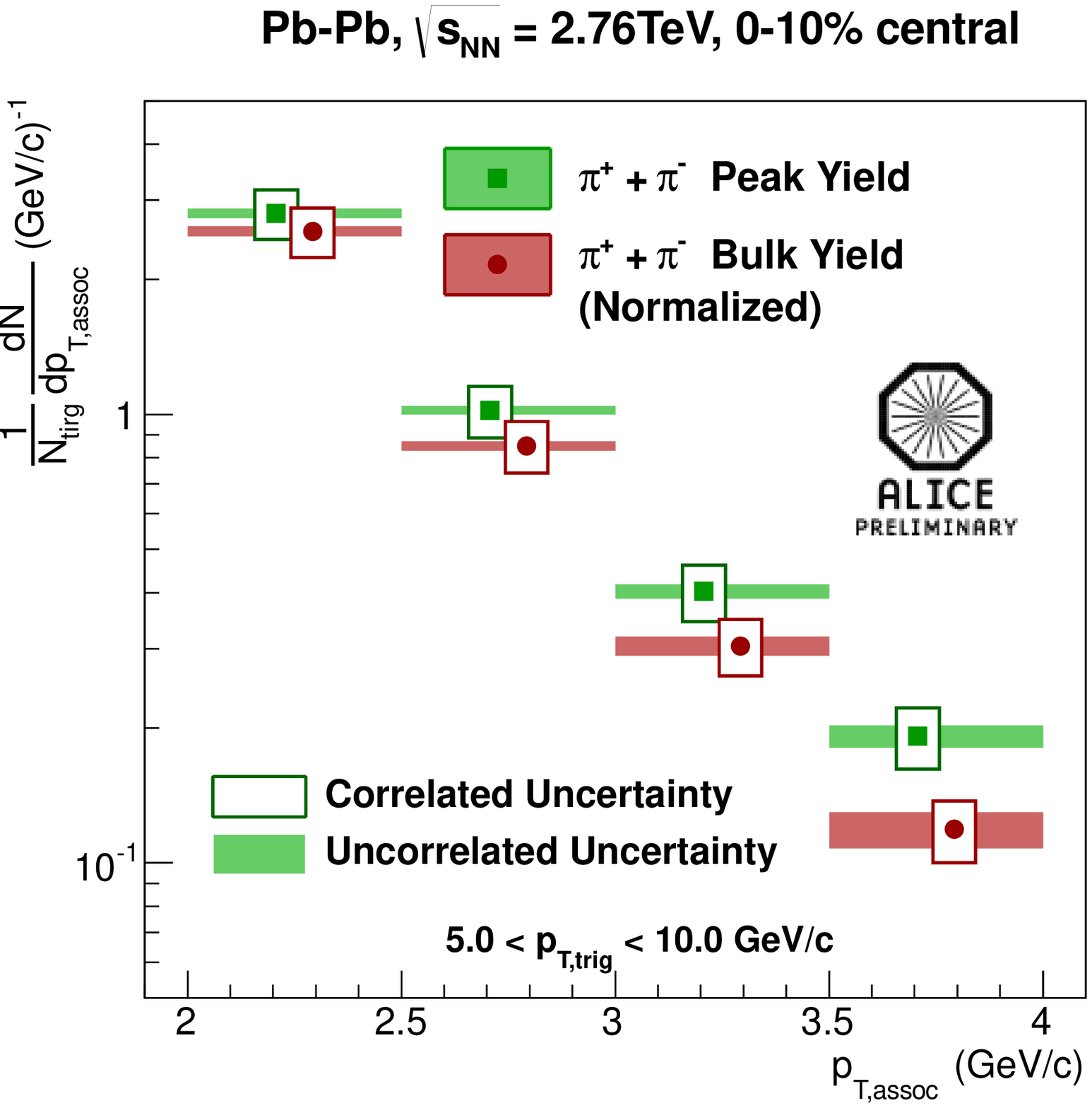}
	\end{subfigure}
	\begin{subfigure}[b]{0.39\textwidth}
		\centering
		\includegraphics[width=\textwidth]{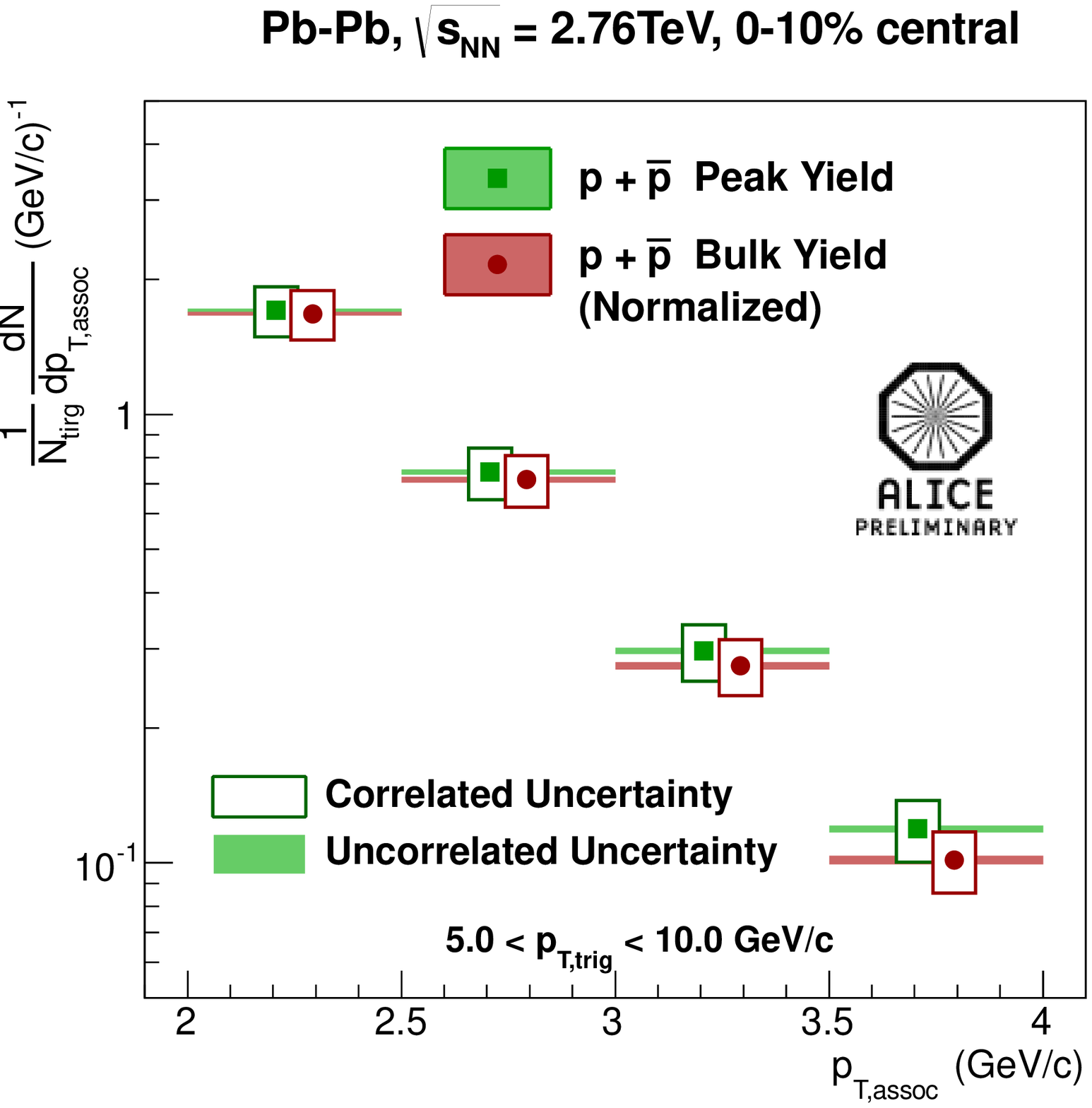}
	\end{subfigure}
	
	\begin{subfigure}[b]{0.39\textwidth}
		\centering
		\includegraphics[width=\textwidth]{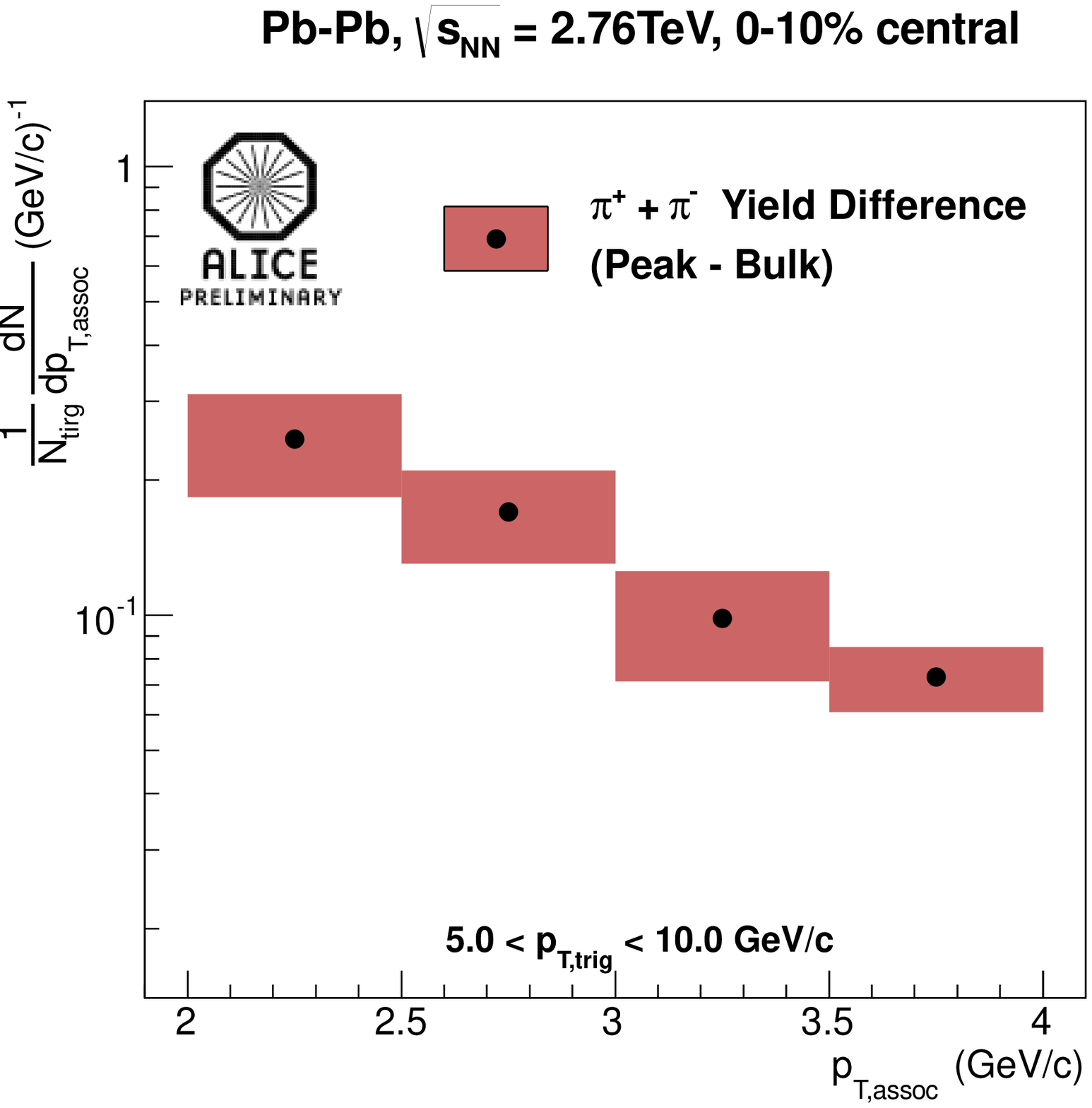}
	\end{subfigure}
	\begin{subfigure}[b]{0.39\textwidth}
		\centering
		\includegraphics[width=\textwidth]{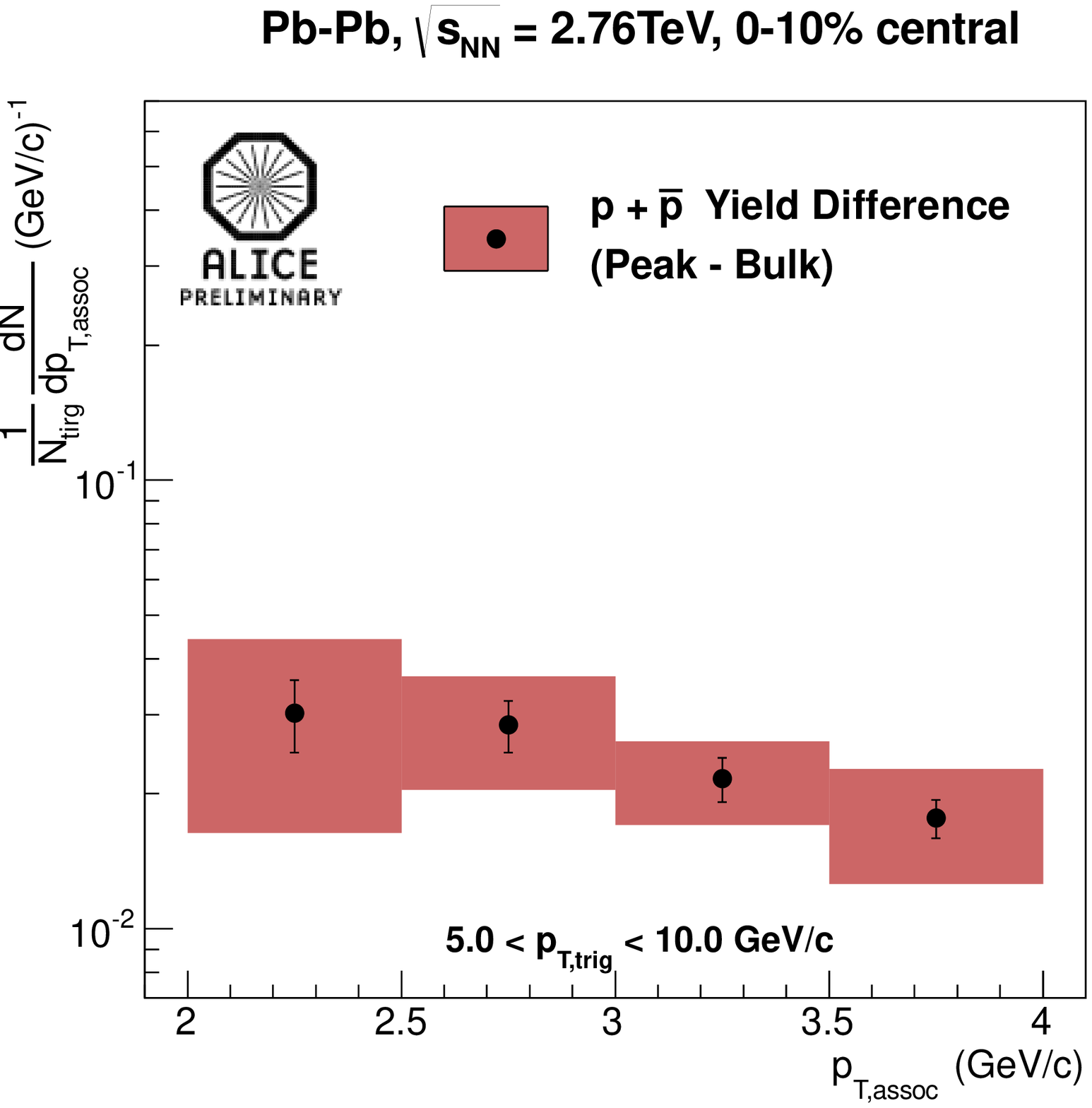}
	\end{subfigure}
	\caption{Top row: pion and proton yield in both peak region and bulk region. The bulk yield is normalized to the area of the peak region in $(\Delta\phi,\Delta\eta)$. Bottom row: jet yield, i.e., peak yield with the bulk subtracted for both protons and pions. Results are not feed down corrected.}\label{fig:Yields}
\end{figure}

On the top row of Fig. \ref{fig:Yields}, the yields per trigger particle in the peak- and bulk region are shown, both for pions and protons (not feed down corrected). Here, the bulk yield is normalized to the area of the peak region, i.e., it has been multiplied by a factor $A_{Peak}/A_{Bulk}$. Correlated and uncorrelated uncertainties are shown separately. In the same figure, on the bottom row, the yields per trigger particle in the jet ($N_{i,Peak}-(A_{Peak}/A_{Bulk})N_{i,Bulk}$) are shown, for $i\in\{\pi,p\}$. 

\begin{SCfigure}[0.5]
	\centering
	\includegraphics[width=0.53\textwidth]{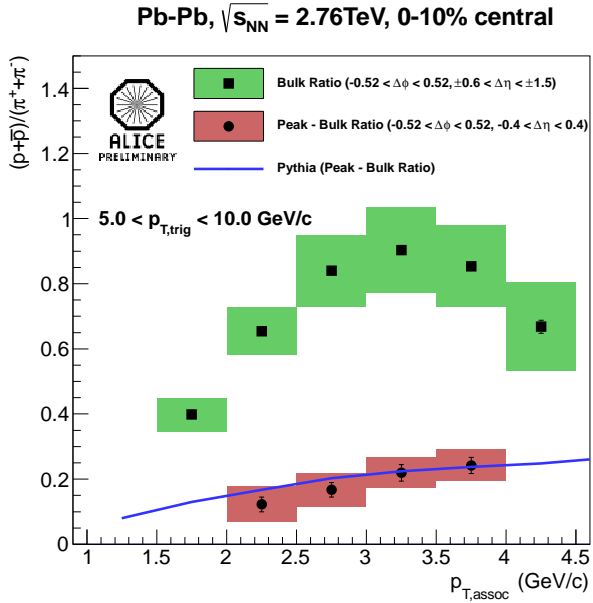}
	\caption{Measured $p/\pi$ ratio in the jet and bulk of a Pb--Pb di-hadron correlation. Comparison with PYTHIA. Results are not feed down corrected.}\label{fig:Ratio}
\end{SCfigure}

Using Eq. (\ref{eq:RatioDef}), the $p/\pi$ ratio was calculated in the bulk and in the jet (see Fig. \ref{fig:Ratio}). For comparison, the jet ratio was calculated using simulated p+p collisions on generator level. For the simulation the default tune of PYTHIA v6.4.21 was used.

The ratio in the bulk increases steeply, up to a value close to unity, and then falls off. This behavior is compatible with inclusive measurements of the $p/\pi$ ratio \cite{IdentifiedHadronsPHENIX2003, IdentifiedHadronsSTAR2010Pub,IdentifiedHadronsALICE2011}. The ratio in the jet ($R^{p/\pi}_{Jet}$) however is compatible with the ratio for jets in PYTHIA, where fragmentation in the vacuum is the dominant mechanism for particle production. 

\section{Conclusion}
A very clear increase of the $p/\pi$ ratio is observed in the bulk of central Pb--Pb collisions compared to the PYTHIA reference. This observation is consistent with earlier inclusive measurements. The $p/\pi$ ratio in the yield associated with a high-$p_T$ trigger particle after background subtraction however is compatible with the PYTHIA reference, suggesting that particle production is unmodified by the presence of the medium. One possible explanation is that fragmentation of energetic partons occurs outside the medium. Further investigation of the $p/\pi$ ratio in the away-side jet may be sensitive to the path length dependance of the effect.





\bibliographystyle{elsarticle-num}
\bibliography{proceedings}






\end{document}